\documentclass[twocolumn,superscriptaddress,showpacs]{revtex4}

\usepackage{amsmath}
\usepackage{amstext}
\usepackage{latexsym}
\usepackage[dvips]{graphicx}

\newcommand{\ket}[1]{\left\vert#1\right\rangle}
\newcommand{\bra}[1]{\left\langle#1\right\vert}

\newcommand{\su}{\uparrow}
\newcommand{\giu}{\downarrow}

\newlength{\defbaselineskip}
\newcommand{\vet}[1]{\mathbf{#1}}

\begin{document}

\title{Berry phase for a spin 1/2 in a classical fluctuating field}
\author{Gabriele De Chiara}
\affiliation{ NEST- INFM \& Scuola Normale Superiore, piazza dei
Cavalieri 7 , I-56126 Pisa, Italy}
\author{G.Massimo Palma}
\affiliation{ NEST- INFM \& Dipartimento di Tecnologie
dell'Informazione, Universita' degli studi di Milano\\ via
Bramante 65, I-26013 Crema(CR), Italy}

\date{\today}
\begin{abstract}
 The effect of fluctuations in the classical control parameters on the Berry phase of a
 spin 1/2 interacting with a adiabatically cyclically varying magnetic field is
 analyzed. It is explicitly shown that in the adiabatic limit dephasing is due
 to fluctuations of the dynamical phase.
\end{abstract}

\maketitle


Berry phase \cite{Berry}  and related geometrical phases
\cite{shapere,zee} have received renewed interest in recent years due
to several proposal for their use in the implementation of quantum
computing gates \cite{ws,Jones,Zanardi,falci00,ekert,cleve,barenco,wilhelm,zoller,keiji,ekertgqc}. Such interest is motivated by the belief that
geometric quantum gates  should exhibit an intrinsic fault
tolerance in the presence of external noise. Such belief is based
on the heuristic argument that being Berry phases geometrical in
their nature, i.e. proportional to the area spanned in parameter
space, any fluctuating perturbation of zero average should indeed
average out. Although this argument seems convincing to the best
of our knowledge it has not been quantitatively probed so far. In
particular although several papers \cite{blais,spiller,gefen} have investigated aspects of
Berry phases in the presence of quantum external noise we are not
aware of any in which  the effect of classical noise in a simple
model of qubit, namely a spin 1/2 interacting with an external
classical field with a fluctuating component has been analyzed.
This is precisely the aim of this paper. For such system the
effects of classical fluctuations in the control parameter on both
geometric and dynamic phases is studied and their impact on
dephasing analyzed.
 Our system consists of a spin 1/2 in the
presence of an external static magnetic field, whose Hamiltonian,
in appropriate units, takes the form
\begin{equation}
H (t)= \frac{1}{2}{\bf B}(t)\cdot \vec{\sigma} \label{hamiltonian1}
\end{equation}
where $\vec{\sigma}= (\sigma_x,\sigma_y,\sigma_z)$, $\sigma_i$ are
the Pauli operators and ${\bf B}(t) = B_0(t) \hat{n}(t)$ with the
unit vector $\hat{n}=(\sin\vartheta\cos\varphi,
\sin\vartheta\sin\varphi, \cos\vartheta)$. The classical field
${\bf B}(t)$ acts as an external control parameter, as its
direction and magnitude can be experimentally changed. When varied
adiabatically the instantaneous energy eigenstates follow the
direction of $\hat{n}$ and therefore can be expressed as
\begin{eqnarray}
|\uparrow_n \rangle &=&e^{-i\varphi/2} \cos \frac{\vartheta }{2}|\uparrow \rangle
+ e^{i\varphi/2}\sin \frac{\vartheta }{2}|\downarrow \rangle\nonumber\\
|\downarrow_n \rangle &=& e^{-i\varphi/2}\sin
\frac{\vartheta}{2}|\uparrow\rangle - e^{i\varphi/2}\cos \frac{\vartheta
}{2}| \downarrow\rangle
\end{eqnarray}
where $|\uparrow \rangle , |\downarrow \rangle$ are the
eigenstates of the $\sigma_z$ operator.

When the time evolution is cyclic i.e. when after a time $T$ we
have ${\bf B}(T) = {\bf B}(0)$ the energy eigenstates acquire a
phase factor which contains a geometric correction to the dynamic
phase:
\begin{equation}
|\uparrow_{n(T)} \rangle = e^{i\delta}e^{i\gamma_B}|\uparrow_{n(0)}
\rangle
\end{equation}
where the dynamic phase $\delta = \int_0^T B_0 (t) dt$ and the
Berry phase can be expressed in term of the so called Berry
connection ${\bf A}^{\uparrow} = i\langle\uparrow_n |
\nabla_{\lambda}|\uparrow_n\rangle$  as $\gamma_B = \oint {\bf
A}^{\uparrow}\cdot d \vec{\lambda}$ where $\vec{\lambda} $ is the
set of control parameters. In our specific example $\vec{\lambda}=
(\vartheta , \varphi)$. An analogous expression holds for ${\bf
A}^{\downarrow}$, relative to the $|\downarrow_{n}\rangle$ state.
It is straightforward to calculate the components of ${\bf A}$:
\begin{eqnarray}
A_{\varphi}^{\uparrow} & = & - A_{\varphi}^{\downarrow} = i\langle \uparrow_n |\partial/\partial \varphi |
\uparrow_n \rangle = \frac{1}{2}\cos\vartheta \\
A_{\vartheta}^{\uparrow} & = & - A_{\vartheta}^{\downarrow} = i\langle \uparrow_n |\partial/\partial \vartheta |
\uparrow_n \rangle = 0
\end{eqnarray}
It is important to note that while the eigenenergies depend on $
B_0(t)$ the eigenstates depend only on $\hat{n}(t)$. As a
consequence the Berry phase depends only on $ \vartheta , \varphi
$. A standard example is a slow precession of ${\bf B}$ at an
angle $\vartheta$ around the $z$ axis with angular velocity
$\omega = 2\pi / T \ll B_0$. A straightforward calculation shows
that

\begin{equation}
\gamma_{\su}  = - \gamma_{\downarrow}= \int_0^{2\pi} A_{\varphi}^{\su} d\varphi =\pi \cos\vartheta
\label{gamma}
\end{equation}

Note that the the Berry phase is proportional to the solid angle subtended by ${\bf B}$ with respect to the degeneracy
${\bf B} = 0$

We are now in the position to extend our analysis to the case in which the magnetic field contains a fluctuating
component. In this case Hamiltonian (\ref{hamiltonian1}) is modified as follows

\begin{equation}
H (t)= \frac{1}{2}\vet{B}_T\cdot \vec{\sigma} =
\frac{1}{2}(\vet{B}(t)+\vet{K}(t))\cdot
\vec{\sigma}\label{hamiltonian2}
\end{equation}

where we have divided the total magnetic field $\vet{B}_T$ into an
average component $\vet{B}$ experimentally under our control and a
fluctuating field $\vet{K}$. We will analyze the case in which
$\vet{B}$ is a field of constant amplitude which undergoes a
cyclic evolution while the components of $\vet{K}$  are random
processes with zero average and small amplitude compared to
$\vet{B}$,  in order to consider lowest order corrections. Finally
we will assume that the fluctuations are characterized by
timescales such that the adiabatic approximation holds. We will
show that this is not an unphysical restriction.

 \begin {figure}[ht]
 \centering
 \includegraphics[scale=0.85]{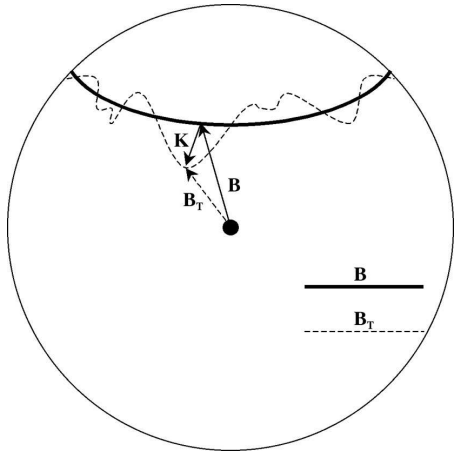}
 \caption {Generic path of the total magnetic field $\vet{B}_T$. The field $\vet{B}$ is under experimental
 control and undergoes a cyclic evolution while $\vet{K}$ is a small random fluctuating field.
 Note that in general $\vet{K}$ fluctuates both in the direction parallel and perpendicular to $\vet{B}$.}
 \label{figure1}
 \end {figure}

Corrections to (\ref{gamma})  have a twofold origin as the
fluctuating field $\bf{K}$ modifies both the connection $\bf{A}$
and the path. To first order correction the connection is
\begin{align}
            A_{\varphi}(\vartheta)&\cong A_{\varphi}(\vartheta_0) +
            \frac{\partial A_{\varphi}}{\partial\vartheta}\delta\vartheta=        \label{E:Aapprox1}\\
    &=\frac{1}{2}\left(\cos\vartheta_0-\delta\vartheta\sin\vartheta_0 \right)     \label{E:Aapprox2}
\end{align}
where $\vartheta_0$ is the polar angle of the average field
$\bf{B}$ while $\delta\vartheta$ is the fluctuation of the polar
angle due to the fluctuating field $\bf{K}$. In order to analyze
the corrections to the path, with no loss of generality,  we again
consider the case of a slow precession of $\bf{B}$ around the
$\hat{z}$ axis. In the presence of $\bf{K}$ the line element
$d\vec{\lambda}$ will have also a component perpendicular to the
orbit of  $\bf{B}$ . However as the connection  $\bf{A}$ has zero
component in the $\vartheta$ direction we can restrict our
attention to the $\varphi$ component of $d\vec{\lambda}$. To this
end we write
\begin{equation} \label{E:dl}
d\lambda=\dot{\varphi}dt\cong (\dot{\varphi}_0 + \delta \dot{\varphi})dt
\end{equation}
In \eqref{E:dl} $\dot{\varphi}_0$ is the average angular velocity.
In our case $\dot{\varphi}_0 =2\pi/T$ while $\delta \dot{\varphi}$
is the first order correction due to $\vet{K}$: in particular when $\vet{K}$
fluctuates in the same direction of $\vet{\dot{B}}$ the precession speed
increases while it decreases in the case of fluctuation in the
opposite direction. Note that fluctuations in the path contain
only corrections in $\varphi$ while fluctuations in the connection
depend only on $\vartheta$. This is independent from any
approximation but is due to the structure of the connection.

We can now express the Berry phase in the presence of noise as
\begin{align}
\label{E:gammapprox3}
\gamma_B &= \int_0^T  (A_{\varphi}(\vartheta_0) + \delta
A_\varphi)(\dot{\varphi}_0 + \delta \dot{\varphi})dt   \\
&\cong  \gamma_B^0   +  \frac{2\pi}{T}\int_0^T \delta
A_\varphi dt  + A_{\varphi}(\vartheta_0) \int_0^T  \delta \dot{\varphi}dt \nonumber\\
& = \gamma_B^0    -   \frac{\pi}{T}\int_0^T
\sin\vartheta_0\delta\vartheta dt            +
A_{\varphi}(\vartheta_0)\delta\varphi(T) \nonumber
\end{align}
where the average Berry phase $\gamma_B^0$  coincides with
$\gamma_B$ in the absence of noise and it has been assumed
$\delta\varphi(0)=0$. The last term in \eqref{E:gammapprox3} is a
non-cyclic contribution which appears when, due to the presence of
$\vet{K}$, $\vet{B}_T$ does not return to its original. In this
case, instead of the geometrical phase definition given by Berry,
which assumes that the Hamiltonian is periodic, we have to use the
definition by Samuel and Bhandari \cite{SamBha88} about non cyclic
evolution. If this is done the third term does not appear and
eq.\eqref{E:gammapprox3} becomes:
\begin{equation}  \label{E:gammapprox4}
\gamma_B = \gamma_B^0  - \frac{\pi}{T}\int_0^T \sin\vartheta_0\delta\vartheta dt
\end{equation}

In order to proceed a physical model for the noise is needed, in
other words  a stochastic process for $\vet{K}$ must be assigned.
Given the probability distribution for the field it is
straightforward to calculate the distribution for the Berry phase.

As a first step we express the trigonometric functions appearing
in \eqref{E:gammapprox4} in terms of the fluctuating field
components $K_i$. This will be useful in calculating the
probability distribution for $\gamma_B$. Let $\vartheta_0$ ,
$\varphi_0$ be the polar angles of $\vet{B}$; $\vartheta$ ,
$\varphi$ those of $\vet{B_T}$; and  $\delta\vartheta$ ,
$\delta\varphi$ the  first order differences between the polar
angles of the two fields. Moreover let $B=|\vet{B}|$ be the
modulus of $\vet{B}$. If we expand in Taylor series
$\cos\vartheta$ we obtain:

\begin{equation} \label{E:cos}
\cos(\vartheta_0+\delta\vartheta)\cong\cos\vartheta_0-\delta\vartheta\sin\vartheta_0=
        \frac{B_3}{B} +
        \frac{K_3}{B}-\frac{B_3}{B^3}\vet{B}\cdot\vet{K}\nonumber
\end{equation}
and therefore
\begin{equation} \label{E:sin}
-\delta\vartheta\sin\vartheta_0=\frac{K_3}{B}-\frac{B_3}{B^3}\vet{B}\cdot\vet{K}
\end{equation}

Substituting equation \eqref{E:sin} in \eqref{E:gammapprox4} we find:
\begin{equation} \label{E:finale}
\gamma_B=\gamma_B^0
+\frac{\pi}{T}\int_0^T\left[\frac{K_3}{B}-\frac{B_3}{B^3}\vet{B}\cdot\vet{K}\right]dt
\end{equation}
From this expression it is possible to find the probability
distribution for $\gamma_B$, once that for $K_i$ is known.

We will assume that the fluctuating field $K_i$ is a
Ornstein-Uhlenbeck process, i.e it is gaussian, stationary and
markovian with a lorentzian spectrum whose  bandwidth $\Gamma_i$ we assume to
be much less than the Bohr frequency of our energy eigenstates.
However while we must have $\omega\ll B$ and $\Gamma_i\ll B$ we have no restriction
on the relative value of $\omega$   vs  $\Gamma_i$. In
order to allow for the possibility of anisotropic noise we assume
that $K_3$ has variance $\sigma_3$ and bandwidth $\Gamma_3$ while
$K_1$ and $K_2$ have $\sigma_{12}$ and $\Gamma_{12}$.

With these assumption we found that the distribution for
$\gamma_B$ is a gaussian whose mean value, as said before, is the
noiseless Berry phase and whose  variance  is
\begin{widetext}
\begin{equation}  \label{E:varBerry}
\sigma^2_\gamma= 2\sigma_{12}^2
                                \left(
                                      \frac{\pi\cos\vartheta_0\sin\vartheta_0}
                                      {TB}
                               \right)^2
                               \left[
                                      \frac{(e^{-\Gamma T}-1)(\Gamma^2-\omega^2)}
                                           { (\Gamma^2+\omega^2)^2} +
                                      \frac{\Gamma T}{\Gamma^2+\omega^2}
                              \right]
            +
                 2\sigma_3^2
                             \left(
                                   \frac{\pi\sin^2\vartheta_0}
                                        {TB}
                             \right)^2
                             \left[
                                   \frac{\Gamma T-1+ e^{-\Gamma T} }
                                        {\Gamma^2}
                             \right]
\end{equation}
\end{widetext}

This expression has an interesting limiting value when
$\Gamma_i\ll\omega$ and $\Gamma_i\gg\omega$. To first order in $\Gamma T$:
\begin{equation}  \label{E:varBerry1}
\!\sigma^2_\gamma\!=       4\sigma_{12}^2\!\!
                                 \left(\!\!
                                       \frac{\pi\!\cos\vartheta_0\!\sin\vartheta_0}
                                            {B}
                                \! \!\right)^2\!\!
                                 \frac{\Gamma_{12} T}
                                      {(2\pi)^2}
               +
                    2\sigma_3^2\!\!
                                \left(\!
                                      \frac{\!\pi\sin^2\!\vartheta_0}
                                           {B}
                              \!  \!\right)^2\!\!\!
                                \left(\!
                                      \frac{1}{2}-\frac{\Gamma_3 T}{6}\!
                                \right)\end{equation}
 We see that the leading term is
$\sigma_3^2\left(\frac{\pi\sin^2\vartheta_0}{B}\right)^2$ which tends
to zero for
little $\vartheta_0$.

 When $T\gg\Gamma^{-1}$ the fluctuating field has time enough to make many uncorrelated
 oscillations during the cyclic evolution. In this case the effect of the fluctuations
 averages out and in the limit $(\Gamma T)^{-1} \to 0$ do not give contribution to the variance which tends to zero:
\begin{equation}  \label{E:varBerry2}
\!\sigma^2_\gamma=  2\sigma_{12}^2\!
                                \left(\!
                                   \frac{\pi\cos\vartheta_0\sin\vartheta_0}{B}\!
                                \right)^2 \!\!\frac{1}{\Gamma_{12} T}
                +
                   2\sigma_3^2
                               \left(
                                    \frac{\pi\sin^2\vartheta_0}{B}
                               \right)^2 \!\! \frac{1}{\Gamma_3 T}
\end{equation}
This is to be  compared to the dynamical phase which grows
linearly in $T$. This different behavior is due  to the fact that while
Berry phase corrections are proportional to $1/T\int K dt$,
corrections to the dynamical phase are proportional to  $\int K
dt$. For a OU process the variance of the integral grows linearly
for times long compared to the autocorrelation time of the field.
This is analogous to the variance of the position of a brownian
particle.

Until now we concentrated only in the geometrical phase. However
during an adiabatic cyclic evolution the eigenstates acquire both
the dynamical and geometrical phase. It is known that the
dynamical phase $\delta$ is proportional to the modulus of the
magnetic field. This means that the dynamical phase becomes a
stochastic processes like the Berry phase. We can write $\delta$
in terms of the fields $\vec{B}$ and $\vec{K}$ as we did for the
geometrical phase:
\begin{equation} \label{dinamica}
\delta= \delta_0 + \int_0^T \frac{\vec{B}\cdot\vec{K}}{B} dt
\end{equation}
where $\delta_0 = B T$ Note that expression \eqref{dinamica} is
similar to \eqref{E:finale} for Berry phase. The difference is
that while $\gamma_B$ comes from an integral in parameter space,
$\delta$ from an integral in the time domain. For instance this
means that if we double time $T$, $\gamma_B$ scales with $T^{-1}$
while the domain of integration of $\delta$ doubles. As we will
see this is crucial for the different role of the two phases in
dephasing.

 Following the same steps as for the Berry phase it is possible
to demonstrate that $\delta$ is a stochastic processes with a
gaussian distribution. Now we analyze the effect of noise on the
coherence of a system, in other words dephasing. Suppose we
prepare the system in a state which is a superposition of the two
eigenstate of the Hamiltonian:
\begin{equation}  \label{E:super}
\ket{\psi}=a\ket{\su}+b\ket{\giu}
\end{equation}
After a slowly cyclic evolution the eigenstates have acquired both
the dynamical and geometrical phases and the final state is:
\begin{equation}  \label{E:super1}
\ket{\psi'}=ae^{i\alpha}\ket{\su}+be^{-i\alpha}\ket{\giu}
\end{equation}
where $\alpha=\gamma_B+\delta$ is the total phase.  In the
presence of noise this phase is a random variable with a gaussian
distribution $P(\alpha)$ then actually the system at the end of
the evolution is in a mixed state.

It can be described by the density operator which is given by the expression:
\begin{equation}       \label{E:rhoalpha}
\rho=\int \ket{\psi'}\bra{\psi'} P(\alpha) d\alpha
\end{equation}
We want to stress that $P(\alpha) \neq P(\gamma_B)P(\delta)$, i.e.
dynamical and geometrical phases are not independent processes
because both depend on $\vec{K}$.

If we insert eq.\eqref{E:super1} in the definition of $\rho$ we find that the population are unchanged while
the coherence are shrunk by a factor $\exp(-2\sigma^2_\alpha)$. In terms of the Bloch vector, this means
that the $z$ component is unchanged while the component parallel to the $xy$ is reduced. This is what is
called dephasing because the relative phase in a superposition is undefined.

In order to do not lose in generality and to compare the dynamical
and geometrical phase we have studied the two case together. We
have found the probability distribution $P(\alpha)$ for $\alpha$
which has mean value $\langle \alpha\rangle = \langle
\gamma_B\rangle + \langle \delta\rangle$ and variance:

\begin{align}
\label{E:varalpha} \sigma_\alpha^2 & = 2\frac{\sigma_{12}^2}{B^2}
       \left( -\frac{\pi\cos\vartheta_0\sin\vartheta_0}{T}  + B \sin\vartheta_0
       \right)^2\times \\
       & \left[ \frac{(e^{-\Gamma_{12} T}-1)(\Gamma_{12}^2-\omega^2)}{ (\Gamma_{12}^2+\omega^2)^2} +
             \frac{\Gamma_{12} T}{\Gamma_{12}^2+\omega^2}     \right]\nonumber \\
& + 2\frac{\sigma_3^2}{B^2}\left( \frac{\pi\sin^2\vartheta_0}{T} +
B \cos\vartheta_0\right)^2
       \left[ \frac{\Gamma_3 T-1+ e^{-\Gamma_3 T}
       }{\Gamma_3^2}\right]\nonumber
\end{align}
In eq. \eqref{E:varalpha} $\sigma_\alpha^2$ is the sum of two
terms one coming from fluctuation in z direction and one in the xy
plane. Each of these terms contains a factor in round brackets in
which we recognize a geometrical term proportional to $1/T$, as we
found in \eqref{E:varBerry} and a dynamical one proportional to
the Bohr frequency. Now because of the adiabaticity condition we
have that the first is much less than the second. As a consequence
the main contribution to dephasing has dynamical rather than
geometrical origin. This does do not mean of course that in a
system in which Berry phase emerges dephasing is less than in a
system in which it does not. What we have demonstrated is that
fluctuations in Berry phase do not contribute considerably to
dephasing.

Another relevant aspect to stress is that our calculation was
performed under the assumptions of a constant circular precession,
however our results are independent of the specific path executed
by the magnetic field as long as the adiabatic approximation is
valid.

It is worth noting that dephasing is  not the only decoherence source in our system.
The Bloch vector in fact does not return to its initial position since the magnetic
field does not. To calculate the correct Bloch vector we have to average the final
positions. However the contributions from this effect are proportional to
$(K_i/B)^2$ and so we can neglect this effect at our level of approximation.

In this paper we have calculated the distribution of Berry phase in the presence of classical noise. Assuming a
OU process for noise we found that under the assumption of small fluctuation Berry phase is a gaussian variable.
We have calculated its mean value and the variance and we found that the variance diminishes as $1/T$. This is
to be compared to the variance of the dynamical phase which grows linearly with $T$. This shows that
adiabaticity and the geometrical aspects of Berry phase reduce fluctuations. This is what was expected but never
demonstrated. Another aspect that makes Berry phase more robust than dynamical phase is that of dephasing. We
showed that geometrical dephasing is much less than the dynamical one. This is due mainly to adiabaticity. This
means that probably quantum gates based upon geometrical phase are more resistant. We would like to point out
that after we submitted our manuscript a paper has appeared \cite{carollo} in which the noise is treated fully
quantum mechanically and similar conclusions have been drawn (but in a completely different setting) about the
invariant of the geometric phase. These two results together complement each other.

\section*{Acknowledgments}
We would like to thank Dr. A.Carollo, Prof.G.Falci, Prof.R.Fazio, Dr.E.Paladino and Dr.V.Vedral for helpful
discussions. This work was supported in part by the EU under grant IST - TOPQIP, "Topological Quantum
Information Processing" Project, (Contract IST-2001-39215).

\end{document}